%% file: charpoly-sigmalu.tex
\RequirePackage{amssymb}
\documentclass[sigconf,natbib=true,screen=true]{acmart}

\usepackage{etoolbox}
\newbool{abstractonly}
\boolfalse{abstractonly} % whole paper
\copyrightyear{} 
\acmYear{} 
\setcopyright{none}
\acmConference[]{}{}{}
\acmBooktitle{}
\acmPrice{}
\acmDOI{}
\acmISBN{}

\usepackage{amsmath,amsthm}
\usepackage{xspace}
\usepackage{algorithm}
\usepackage[noend]{algpseudocode} %%% this is for the algorithms
\algnewcommand{\algorithmicassumption}{\textbf{Requirement:}}

\usepackage{paralist}
\usepackage{multirow}
\usepackage[capitalize]{cleveref}

\newcommand{\Z}[1]{\ensuremath{{Z}_{{#1}}}\xspace}
\newcommand{\K}{\ensuremath{\mathbb{K}}\xspace}
\newcommand{\I}[1]{\ensuremath{{I}_{{#1}}}\xspace}
\newcommand{\J}[1]{\ensuremath{{J}_{{#1}}}\xspace}
\newcommand{\UU}[1]{\ensuremath{{U}_{{#1}}}\xspace}

\newcommand{\N}{\ensuremath{\mathbb{N}}}

\newcommand{\SO}[1]{\ensuremath{{\tilde{O}\left({#1}\right)}}\xspace}
\newcommand{\GO}[1]{\ensuremath{{O\mathopen{}\left({#1}\right)\mathclose{}}}\xspace}
\newcommand{\GOM}[1]{\ensuremath{{\Omega\mathopen{}\left({#1}\right)\mathclose{}}}\xspace}
\newcommand{\assign}{\ensuremath{\leftarrow\xspace}}
\newcommand{\trsp}[1]{#1^\mathsf{T}}
\newcommand{\SLU}{\texorpdfstring{$\Sigma LU$}{Sigma LU}}
\usepackage[colorinlistoftodos]{todonotes}

\definecolor{ggreen}{RGB}{34,139,34}

\newcommand{\bsgscost}{\ensuremath{n^{\omega - \frac{\omega-1}{5-\omega}}\alpha^{\frac{\omega-1}{5-\omega}}}\xspace}
\title{Computing the Characteristic Polynomial of Generic Toeplitz-like and Hankel-like Matrices}
\author{Pierre Karpman}
\affiliation{%
  \institution{Universit\'e Grenoble Alpes}
  \department{Laboratoire Jean Kuntzmann, CNRS UMR 5224}
  \streetaddress{700 avenue centrale, IMAG --- CS 40700}
  \city{Grenoble}
  \country{France}
}
\author{Cl\'ement Pernet}
\affiliation{
  \institution{Universit\'e Grenoble Alpes}
  \department{Laboratoire Jean Kuntzmann, CNRS UMR 5224}
  \streetaddress{700 avenue centrale, IMAG --- CS 40700}
  \city{Grenoble}
  \country{France}
}
\author{Hippolyte Signargout}
\affiliation{
\institution{Univ. Lyon, ENS de Lyon, CNRS, Inria, UCBL}
  \department{LIP UMR 5668 and LJK UMR 5224}
   \city{Lyon}
   \country{France}
}
\author{Gilles Villard}
\affiliation{
  \institution{Univ. Lyon, CNRS, ENS de Lyon, Inria, UCBL}
  \department{LIP UMR 5668}
  \city{Lyon}
  \country{France}
}
\makeatletter
\hypersetup{%
	pdftitle={\@title},
	pdfauthor={\@author},
	breaklinks=true,
	colorlinks=true,
	citecolor=blue,
}
\makeatother

\begin{document}
\begin{abstract}
New algorithms are presented for computing annihilating polynomials of 
Toeplitz, Hankel, and more generally Toeplitz+ Hankel-like matrices over a field. Our approach
follows works on Coppersmith's block Wiedemann method with structured projections, which have
been recently successfully applied for computing the bivariate resultant. A first baby-step/giant
step approach---directly derived using known techniques on structured matrices---gives a randomized Monte Carlo algorithm for the minimal polynomial of an
\(n\times n\) Toeplitz or Hankel-like matrix of displacement rank \(\alpha\) using 
$\SO{n^{\omega - c(\omega)} \alpha ^{c(\omega)}}$ arithmetic operations, where $\omega$ is the exponent
of matrix multiplication and
 $c(2.373)\approx 0.523$ for the best known value of~\(\omega\).
For generic Toeplitz+Hankel-like matrices a second algorithm computes 
the characteristic polynomial in \(\SO{n^{2-1/\omega}}\)
operations when the displacement rank is considered constant.  Previous algorithms required
$O\left(n^2\right)$ operations while the exponents presented here are respectively less than $1.86$ and $1.58$
with the best known estimate for $\omega$.
%$\tilde O \left(n^{1+\frac{\omega(\omega - 1)}{2\omega -1}}\right)$
\end{abstract}

\keywords{
  Characteristic polynomial, minimal polynomial,
  Toeplitz matrix,
  Hankel matrix,
  Toeplitz+Hankel-like matrix.
}
%-------------------------------------------------------------------------------
\maketitle

%%%% Should we only generate the abstract ?
\ifbool{abstractonly}{
\end{document}
}

%-------------------------------------------------------------------------------
%%%%%%%%%%%%%%
%%%%%%%%%%%%%%
\section{Introduction}\label{sec:intro}

We consider the problem of computing the minimal or the characteristic polynomial of 
Toeplitz-like and Hankel-like matrices, which include Toeplitz and Hankel ones. The necessary definitions about those structures are given in \cref{sec:prelim:rankdisp}.

Throughout the paper $T\in \K^{n\times n}$ is non-singular %% \gvo{Faire \'eventuellement un shift al\'atoire au d\'epart.}
and 
either Toeplitz-like or Hankel-like, where $\K$ is a commutative field. The structure is parameterized by the displacement rank $1 \leq \alpha \leq n$ of $T$~\cite{KKM79,pan01}.
In particular a Toeplitz or a Hankel matrix has displacement rank $\alpha=2$. 

The determinant of $T$ can be computed in $\SO{\alpha ^{\omega -1}n}$ operations in $\K$, where $\omega \leq 3$ is a feasible exponent for square $n\times n$ matrix multiplication. For the best known value of $\omega$ 
one can take 
$\omega \approx 2.373$~\cite{LeGall14,alman2020refined}. When $T$ has generic rank profile (the leading principal submatrices are non singular) a complexity bound 
$\SO{\alpha ^{2}n}$ for the determinant is derived from~\cite[Cor.~5.3.3, p. 161]{pan01}.
In the general case, for ensuring the rank profile one uses rank-regularization techniques initially developed in~\cite{KaSa91,Kal94} that lead to randomized Las Vegas algorithms assuming that the cardinality of $\K$ is large enough; 
see~\cite[Sec. 5.6-5.7]{pan01} and \cite{BJMS17} for detailed studies in our context.
Taking advantage of fast matrix multiplication is possible using the  results in
\cite{BJMS17},
where fundamental matrix operations, including the determinant, are performed in time $\SO{\alpha ^{\omega -1}n}$
for a wide spectrum of displacement structures. In this approach the determinant is revealed by the recursive
  factorization of the inverse.

The characteristic polynomial  $\det (x\I{n} - T)$ of $T$ is a polynomial of degree $n$. Using an evaluation-interpolation scheme it follows that it can be computed in $\SO{\alpha ^{\omega -1}n^2}$ operations in 
$\K$. We also refer to \cite[Ch. 7]{pan01} for a Newton-Structured iteration scheme in time $\SO{\alpha ^{2}n^2}$. 

For a Toeplitz or Hankel matrix these  complexity bounds 
for computing the characteristic polynomial were quadratic; 
our contribution establishes an 
improved bound $\SO{n^{2-1/\omega}}$ for generic matrices (given in compressed form), which is
sub-quadratic including when using $\omega =3$.
We build on the results of~\cite{Vil18} where only the case of a Sylvester matrix was 
treated, and show that the approach can be generalized to larger displacement rank families. In particular, the Hankel-(like)  
case requires the use of sophisticated techniques in order to handle the 
Toeplitz+Hankel structure~\cite{heinig1988fast,HeRo04} and its generalizations~\cite{pan01}.

%% Starting point the analysis of Coppersmith's
%% block Wiedemann algorithm in~\cite{KaVi05}.

  The   algorithms we propose fit into the broad family of Coppersmith's block Wiedemann algorithms ; we refer to~\cite{KaVi05} for the 
  necessary material and detailed considerations on the approach. Another interpretation in terms of structured lifting and matrix fraction reconstruction is given in~\cite{Vil18}. 

From $T\in \K^{n\times n}$, the problem is to compute the determinant (or a divisor) 
of the characteristic matrix $M(x)=x\I{n}-T$. For $1\leq m \leq n$ and well chosen projection 
matrices $V$ and $W$ in $\K^{n\times m}$, the principle is to reconstruct an irreducible fraction description 
$P(x)Q^{-1}(x)$ of $\trsp{V}M(x)^{-1}W \in \K(x)^{m\times m}$, where $P, Q \in \K[x]^{m \times m}$, from a truncated series expansion of the fraction.
The denominator matrix $Q$ carries information on the Smith normal form of $M(x)$~\cite[Thm. 2.12]{KaVi05}. 
Using random $V$ and $W$ allows to recover the minimal polynomial of $T$ from the largest invariant factor of 
$M(x)$, and for a generic matrix~$T$ the characteristic polynomial is obtained~\cite{KaVi05,Vil18}.

The matrix $Q$ is computed from a truncation $S^{(m)}\in \K[x]^{m\times m}$ of the series expansion of $\trsp{V}M(x)^{-1}W$,
\begin{equation} \label{eq:defseries}
S^{(m)}(x) =  - \sum_{k\geq 0}^{2 \lceil n/m \rceil} \trsp{V}(T^{-1})^kW x ^k
\end{equation}
using for example matrix fraction reconstruction~\cite{BeLa94,GJV03}. We will not detail 
these latter aspects in this paper since they can be found elsewhere in the literature: see~\cite{KaVi05,Vil18} for the general techniques involved; \cite[Cor. 6.4]{Vil97:TR} for the power series truncation; and \cite{KaYu13} for alternative 
reconstruction possibilities. The results we need on matrix polynomials are recalled in \cref{sec:matpoly}.

We focus on the computation of the power series terms $H_k = \trsp{V}(T^{-1})^kW$
in \cref{eq:defseries}.
The idea  for improving the complexity bounds is to use 
 structured
projections $V$ and $W$ in order to speed up the computation of the expansion such as in~\cite{EGGSV07,Vil18}. 
A typical choice is such that the matrix product by $V$ and $W$ is reduced. 
The central difficulty is to show that the algorithm remains correct; special choices for 
$V$ and $W$ could prevent a fraction reconstruction with appropriate cost, or give a denominator matrix 
$Q$ with too little information on the invariant structure of $T$. 

For a generic input matrix and our best exponent, in \cref{sec:compressed}
we follow the choice of~\cite{Vil18} and work with $V=W=X$ where  
\(X=\trsp{\begin{pmatrix}  \I{m}&0\end{pmatrix}} \in \K^{n\times m}\). 
An $n \times n$ Toeplitz or a Hankel matrix is defined by $2n-1$ elements of $\mathbb{K}$, and
our algorithm is correct except on a certain hypersurface of $\K^{2n-1}$. The same way, a Toeplitz-like or Hankel-like matrix of displacement rank \(\alpha\) is defined by the \(2n\alpha\) coefficients of its generators, and our algorithm is correct for all values of \(\K^{2n\alpha}\) except for a hypersurface.
If $T$ is Hankel, the matrix $M(x)= x\I{n} - T$ is Toeplitz+Hankel and the algorithm involves 
a compressed form that generalizes the use of generators associated to displacement operators~\cite{HeRo04,pan01}. 
The algorithm computes a compressed representation of \(M(x)^{-1}\) modulo \(x^{2\lceil n/m \rceil+1}\), and exploits 
its structure to truncate it into a compressed representation of \(S^{(m)}(x) = \trsp{X}M(x)^{-1}X \mod
x^{2\lceil n/m \rceil+1}\) at no cost. The parameter $m$ can be optimised to get an algorithm using \SO{n^{2 - 1/\omega}} operations when the displacement rank is considered constant.

Before considering the fast algorithm for the generic case, in \cref{sec:bsgs} we consider the baby steps/giant steps algorithm of~\cite{KaVi05}. Indeed, thanks to the incorporation of fast matrix multiplication in 
basis structured matrix operations~\cite{BJMS17}, the overall approach with dense projections $V$ and $W$ 
already allows a slight exponent improvement. Taking into account that the input matrix $T$ is structured, a direct cost analysis of the algorithm of \cite{KaVi05} improves on the quadratic cost for Toeplitz and Hankel 
matrices as soon as one takes $\omega < 3$. However it is unclear to us how to compute the characteristic 
polynomial in this case (see the related Open Problem~3 in \cite{Kal00-2}). The algorithm we propose is 
randomized Monte Carlo and we compute the minimal polynomial in \SO{n^{\omega - c(\omega)}} operations with \(c(\omega) = \frac{\omega - 1}{5-\omega}\). For Toeplitz-like and Hankel-like matrices with displacement rank \(\alpha\), the cost is multiplied by \SO{\alpha^{c(\omega)}}.

\begin{paragraph}{Notation}
Indices of matrix and vectors start from zero.
The vectors of the \(n\)-dimensional canonical basis are denoted by \(e_0^n,\ldots,e_{n-1}^n\).
For a matrix \(M\), \(M_{i,j}\) denotes the coefficient \((i,j)\) of this
matrix, \(M_{i,*}\), its row of index \(i\) and \(M_{*,j}\) its column of index \(j\).
\end{paragraph}
%-------------------------------------------------------------------------------
%%%%%%%%%%%%%%
\input{prelim}
%-------------------------------------------------------------------------------
%%%%%%%%%%%%%%
\input{bsgs}
\input{sigmalu}
%%%%%%%%%%%%%%
\section{Special matrices for genericity}

The generic matrices \(T\) for which our algorithms output the characteristic polynomial are matrices such 
that \(\mathcal H^{(n)}=\mathcal{H}_{1..n,1..n}\) is non-singular (\cref{cor:slu,cor:slu:hank}), where
$$\mathcal H = \left(\trsp VT^{i+j}W\right)_{0\leq i,j \leq \lceil n/m\rceil - 1}$$
The first algorithm is Monte Carlo with  matrices \(V\) and \( W\) sampled at random. In the second
algorithm, however \(V =W =X\) are fixed, and \(\det{\mathcal H^{(n)}}\) is a polynomial in the
coefficients of \(T\). Toeplitz and Hankel matrices have \(2n-1\) independant coefficients. The
coefficients of a Toeplitz-like or Hankel-like matrix of displacement rank \(\alpha\) are themselves polynomials in the coefficients of its generators, so \(\det{\mathcal H^{(n)}}\) is by composition a polynomial on the \(2n\alpha\) coefficients of the \(n\times\alpha\) generators of \(T\).

In this section, we show that \(\det{\mathcal H^{(n)}}\) is not uniformly zero on the space of Toeplitz (resp. Hankel) matrices by finding one Toeplitz (resp. Hankel) matrix for which \(\mathcal H^{(n)}\) is non-singular.
This shows the algorithm is correct for all matrices of each class except for those with coefficients in a certain
variety of \(\K^{2n-1}\).
As the displacement rank of the matrices we show is 2 or less, they are Toeplitz-like (resp. Hankel-like) and can be represented with larger generators (padded with zeros).
The algortihm is thus also correct for matrices with displacement rank \(\alpha \geq 2\) whose generators'
coefficients are not in a certain variety of \(\K^{2n\alpha}\).
Both matrices are also Toeplitz+Hankel and Toeplitz+Hankel-like so the same reasoning shows the algorithm is
correct for all Toeplitz+Hankel matrices except for those with coefficients in a certain hypersurface of
\(\K^{4n-2}\) and all Toeplitz+Hankel-like matrices with displacement rank \(\alpha \geq 4\) except for those on a
certain hypersurface of \(\K^{2n\alpha}\).

\subsection{A Toeplitz Point}\label{sec:point:toep}
%In this section $X=Y=\trsp{[\I{m}\ 0]}$.
Let
\[
T = \left(\begin{array}{cc} 0 & \I{m} \\ -\I{n - m} & 0 \end{array}\right)
\] % \in \K^{n \times n}$
and \(M(x) = x\I{n} - T\). % \in\K^{n \times n}$.
Let $P(x)\in \K[x]^{n \times m}$ defined by:
\begin{align*}
 &P_{n-m+k,k} = 1 &\text{ for }& 0\leq k \leq m \\
 &P_{i,k} = xP_{i+m,k} &\text{ for }& 0\leq k \leq m, 0\leq i \leq n - m - 1
\end{align*}
With
\[
D(x) = \left(\begin{array}{cc}
  0 & x^{\lfloor n/m\rfloor}\I{n \text{ mod } m} \\ x^{\lfloor n/m\rfloor-1}\I{-n \text{ mod } m} &
  0 \end{array}\right)
\]
we can write \(P(x) = \trsp{ \left(\trsp{ D(x)} \quad  {R(x)} \quad {\I{m}}\right)}\). From there we have $M(x)P(x) = \trsp{ \left( x\trsp {D(x)}-\I{m} \quad 0 \right)}$ and thus
  \[
  \trsp XM^{-1}(x)X = \trsp XP(x)\left(xD(x) - \I{m}\right)^{-1}.
  \]
  That is $\trsp XM^{-1}(x)X = D(x)Q^{-1}(x)$ with $Q(x) = xD(x) - \I{m}$. As $x\I{m}D(x) - \I{m}Q(x) = \I{m}$, the fraction $DQ^{-1}$ is irreducible and
    \[
    \det Q = \pm \, x ^{\lfloor n/m\rfloor (n\mod{m}) + (\lfloor n/m \rfloor-1)(-n \mod{m})} - 1
    \]
     from which we get $\deg \det Q = n$. By \cite[Lemma 2.4]{Vil18}, the matrix $\mathcal H^{(n)}$
     is therefore non-singular.

  \subsection{A Hankel Point}\label{sec:point:hank}

  Let \(T_n = (\I{n}+\Z{n}^m)\J{n}\). For
\( j\) such that \(2j \leq \lceil n/m \rceil - 1\),
%\(0\leq j< n/2m\),
rows \(jm\) to \((j+1)m-1\) of \(T_n^{2j}X\) are \I{m} and the following rows are \(0\). This can be seen by recursively applying the band matrix \(T_n^2 = \Z{n}^m + \I{n} + \Z{n}^m\Z{n}^{m\mathsf T} +\Z{n}^{m\mathsf T}\) to \(X\). By applying \(T_n\) to \(T_n^{2j}X\) we get that the rows \(n - (j+1)m\) to \(n - jm -1\) of \(T_n^{2j + 1}X\) are \J{m}, and the preceding rows are \(0\).

Let \(K_r\) be the first $n$ columns of \(\left(T^0X|\hdots|T^{\lceil n/m \rceil-1}X\right)\).
%K_r = \left(T^0X|\hdots|T^{\lceil n/m \rceil-2}X|\left(T^{\lceil n/m \rceil-1}\right)_{*,0\hdots n-1 -m(\lceil n/m \rceil-2)}X\right)
\(K_r\) is non-singular, as its
columns can be permuted to get a matrix of the form
\[
\left(\begin{array}{cc}
  \trsp{L_1} & 0\\
  0 & L_2
\end{array}\right)
\]
where \(L_1\) and \(L_2\) are lower triangular with ones on the diagonal.
As T is symmetric, \(K_l\) defined as the first \(n\) rows of
\[
\trsp{\left(T^{0 \mathsf T}X|\hdots|T^{(\lceil n/m\rceil - 1)\mathsf T}X\right)}
\]
is also non-singular, as well as \(\mathcal H^{(n)} = K_lK_r\).

%The matrix $\J{n}$ is a non-zero point for $\det H$ in the class of Hankel matrices and Hankel-like matrices of displacement rank $1$. Let $M(x) = x\I{n} - \J{n} \in\K^{n \times n}$ and $P(x) = \left(\begin{array}{c}x\I{m} \\ 0 \\ \J{m} \end{array} \right) \in \K[x]^{n \times m}$.

 % We have $M(x)P(x) = \left(\begin{array}{c} (x^2-1)\I{m} \\ 0    \end{array}\right)$ and $\trsp XM^{-1}(x)X = \trsp XP(x)\left((x^2-1) \I{m}\right)^{-1}$. That is $\trsp XM^{-1}(x)X = x\I{m}Q^{-1}(x)$ with $Q(x) = (x^2-1) \I{m}$. As $x\I{m}x\I{m} - \I{m}Q(x) = \I{m}$, the fraction $x\I{m}Q^{-1}$ is irreducible and $\det Q = (x^2-1)^m$, from which we get $\deg \det Q = 2m$\footnote{Do we need exactly $n$?}.

 %As we can write $\trsp XM^{-1}(x)X = \sum \trsp XT^iXx^{-i}$, the minimal generating polynomial of the sequence $(\trsp XT^{i}X)_{i\geq0}$ is of degree $n$.

  %The matrix $H = \left(\trsp XT^{i + j}X\right)_{0 \leq i, j \leq n-1}$ is non-singular.

%-------------------------------------------------------------------------------
%%%%%%%%%%%%%%
%%%%%%%%%%%%%%
%\input{onesidedproj} 

%%%%%%%%%%%%%%%%%%%%%%%%%%%%%%%%%%%%%%%%%%%%%%%%%%%%%%%%%%%%%%%%%%%%%%%%%%%%%%%%
%\appendix
%\section{Proof}

%\input{annexes}

\bibliographystyle{ACM-Reference-Format}

%GV todo, another mean?   bibtool -- 'add.field{title2="\href{http://dx.doi.org/%s(doi)}{%s(title)}"}' biblio.bib -o newbiblio.bib
%\gv{See the comment in the latex, one way or another for having the hyperlink with the title with ACM format.}

\bibliography{biblio}

\end{document}

%% file: prelim.tex
%!TEX root = charpoly-sigmalu.tex

%%%%%%%%%%%%%%
\section{Material for rank displacement structures}\label{sec:prelim:rankdisp}

A wide range of structured matrices are efficiently described by the action of a displacement
operator~\cite{KKM79}.
There are two types of such operators: the Sylvester operators of the form
\[\nabla_{M,N}: A \mapsto MA - AN, \]
and the Stein operators of the form
\[\Delta_{M,N}: A \mapsto A - MAN; \]
where \(M\) and \(N\) are fixed matrices.
A Toeplitz matrix \(T\) is defined by \(2n-1\) coefficients \(t_{-n+1},\dots,t_{n-1} \in \K\) such
that \(T=(t_{i-j})_{i,j}\). Its image through \(\Delta_{\Z{n},\trsp{\Z{n}}}\), where \(\Z{n}=
%\begin{smatrix}
%  0&\\\
%  1&0\\
%  & \ddots&\ddots\\
%  &&1&0
%\end{smatrix}
(\delta_{i, j+1})_{0\leq i,j \leq n-1}\) has rank 2.
Similarly, a Hankel matrix \(H\) is defined by \(2n-1\) coefficients \(h_0,\dots,h_{2n-2}\) such that
\(H=(h_{i+j})_{i,j}\) and its image through  \(\nabla_{\Z{n},\trsp{\Z{n}}}\) has rank 2.

As a generalization, the class of Toeplitz-like (resp. Hankel-like) matrices is
defined~\cite{HeRo84,pan01} as those matrices which image through \(\Delta_{\Z{n},\trsp{\Z{n}}}\) (resp. \(\nabla_{\Z{n},\trsp{\Z{n}}}\)) has a bounded rank \(\alpha\),
called the displacement rank.
%\todo[inline]{Attention pour certains, la def est avec $Z_\alpha$}
Lastly, any sum of a Toeplitz and a Hankel matrix, (forming the class of Toeplitz+Hankel
matrices) has an image of rank 4 through the displacement operator %\(\nabla_{Z+\trsp{Z},Z+\trsp{Z}}\).
\(\nabla_{\UU{n},\UU{n}}\) where $\UU{n} = \Z{n}+ \trsp {\Z{n}}$.
However, contrarily to the previous instances, this operator is no longer regular, and the low rank
image does not suffice to uniquely reconstruct the initial matrix: additional data (usually a
first or a last column) is required for a unique reconstruction.
The class of Toeplitz+Hankel-like matrices is formed by those matrices whose image through
\(\nabla_{\UU{n},\UU{n}}\) has a bounded rank.

\subsection{Product of Structured Matrices}
\begin{proposition}[{\cite[Theorem~1.2]{BJMS17}}]\label{prop:mbv}
   Let \(A\in \K^{n\times n}\) be a Toeplitz-like or Hankel-like matrix with displacement rank \(\alpha\) given by its generators and
   \(B\in\K^{n\times m}\) be a dense matrix. The multiplication  of $A$ by $B$ can be computed in
   \(\SO{n\max(\alpha,m)\min(\alpha,m)^{\omega-2}}\) operations in~\K.
\end{proposition}
\begin{proposition}%[{\cite[Propositon~4]{BJS08}}]
  \label{prop:TxT}
  Let \(A, B\in \K^{n\times n}\) be two Toeplitz-like matrices of
displacement rank \(\alpha\) and \(\beta\)
  respectively, then their product \(AB\) is a Toeplitz-like matrix of displacement rank at most
  \(\alpha+\beta+1\). Furthermore, given generators for \(A\) and \(B\) w.r.t. \(\Delta_{\Z{n},\trsp{\Z{n}}}\), one
    can compute generators for \(AB\) w.r.t. the same operator in
  \(\SO{n(\alpha+\beta)^{\omega-1}}\) field operations.
\end{proposition}
\begin{proof}
Let \(G_A,H_A\) and \(G_B,H_B\) be the generators of \(A\) and \(B\) respectively. They satisfy  A-\(\Z{n}A\trsp{\Z{n}} = G_A \trsp{H_A}\)
and  \(B-\Z{n}B\trsp{\Z{n}} = G_B \trsp{H_B}\).
Consequently
\begin{align*}
AB &=& (\Z{n}A\trsp{\Z{n}} + G_A\trsp{H_A})(\Z{n}B\trsp{\Z{n}} + G_B\trsp{H_B} )\\
  &=& \Z{n}AB\trsp{Z}_n - \Z{n} A_{*,n}B_{n,*}\trsp{\Z{n}} + (\Z{n}A\trsp{\Z{n}}G_B)\trsp{H_B} \\
  &&+ G_A (\trsp{H_A}\Z{n}B\trsp{\Z{n}} + \trsp{H_A}G_B\trsp{H_B}),
\end{align*}
and therefore \(AB-\Z{n}AB\trsp{\Z{n}} = G_{AB}\trsp{H_{AB}}\) for
\begin{align*}
G_{AB} &=& \left(
\begin{array}{c|c|c}
  G_A &  \Z{n}A\trsp{\Z{n}}G_B &  -\Z{n}A_{*,n}
\end{array}
\right)
\\
H_{AB} &=& \left(
\begin{array}{c|c|c}
\Z{n}\trsp{B}   \trsp{\Z{n}}H_A + H_B\trsp{G_B}H_A &  H_B & \Z{n} \trsp{B_{n,*}}
  \end{array}\right),
\end{align*}
thus showing that \(AB\) has displacement rank at most \(\alpha+\beta+1\).

Computing these generators involves applying \(A\) on a dense \(n\times \beta\) matrix and \(B\) on
a dense \(\alpha\times n\) matrix, and computing the product of an \(\alpha \times n\) by an \(n
\times \beta\) matrix and the product of an \(\alpha \times \beta\) by a \(\beta\times n\)
matrix. Using~\cite[Theorem~1.2]{BJMS17}, these cost \SO{n(\alpha+\beta)^{\omega-1}} field
operations.
\end{proof}

\begin{proposition}\label{prop:HxH}
  Let \(A, B \in \K^{n\times n}\) be two Hankel-like matrices of displacement rank \(\alpha\) and \(\beta\)
  respectively, then their product \(AB\) is a Toeplitz-like matrix of displacement rank at most
  \(\alpha+\beta+1\). Furthermore, given generators for \(A\) and \(B\) w.r.t. \(\nabla_{\Z{n},\trsp{\Z{n}}}\),
  generators for \(AB\) w.r.t. \(\Delta_{\Z{n},\trsp{\Z{n}}}\) can be computed in \(\SO{n(\alpha+\beta)^{\omega-1}}\).
\end{proposition}
\begin{proof}
Let \(G_A,H_A\) and \(G_B,H_B\) be the generators of \(A\) and \(B\) respectively, satisfying
\(\Z{n}A-A\trsp{\Z{n}} = G_A \trsp{H_A}\) and \(\Z{n}B-B\trsp{\Z{n}} = G_B \trsp{H_B}\).
Using a similar reasoning as for~\cref{prop:TxT} we can deduce that
%% Consequently
%% \begin{align*}
%%   \Z{n}AB\trsp{\Z{n}} &=& (G_A\trsp{H_A} + A\trsp{\Z{n}})(\Z{n}B- G_B\trsp{H_B} )\\
%%   &=& AB - A_{*,n}B_{n,*} + G_A (\trsp{H_A}\Z{n}B-\trsp{H_A}G_B\trsp{H_B})\\
%%   && - (A\trsp{\Z{n}}G_B)\trsp{H_B},
%% \end{align*}
%% and therefore
\(AB-\Z{n}AB\trsp{\Z{n}} = G_{AB}\trsp{H_{AB}}\) for
\begin{align*}
G_{AB} &=& \left(
\begin{array}{c|c|c}
  G_A &  A\trsp{\Z{n}}G_B &  A_{*,n}
\end{array}
\right)
\\
H_{AB} &=& \left(
\begin{array}{c|c|c}
   {H_B}\trsp{G_B}{H_A} - \trsp{B}\trsp{\Z{n}}H_A &  H_B &  \trsp{B_{n,*}}
  \end{array}\right),
\end{align*}
thus showing that \(AB\) has displacement rank at most \(\alpha+\beta+1\).
Computing these generators again
%% involve applying \(A\) on a dense \(n\times \beta\) matrix and \(B\) on
%% a dense \(\alpha\times n\) matrix, and computing the product of an \(\alpha \times n\) by an \(n
%% \times \beta\) matrix and the product of an \(\alpha \times \beta\) by a \(\beta\times n\)
%% matrix. Using~\cite[Theorem~1.2]{BJMS17}, these operations are
costs  \SO{n(\alpha+\beta)^{\omega-1}} field
operations.
\end{proof}

\begin{proposition}\label{prop:power}
    Let \(A\in \K^{n\times n}\) be a Toeplitz-like (resp. Hankel-like) matrix of displacement rank \(\alpha\),
    then for an arbitrary (resp. even) $r$, \(A^r\) is a Toeplitz-like matrix of displacement rank at most \((\alpha+1) r\)
 and its generators can be computed in \(\SO{n(\alpha r)^{\omega-1}}\) field operations.
\end{proposition}
\begin{proof}
    Using fast exponentiation one computes \(A^r\) as:
  \[
  A^r = \prod\limits_{k=0}^{\lfloor\log r\rfloor}\left(A^{2^k}\right)^{l_k} \text{ where } r = \sum\limits_{k=0}^{\log r}l_k2^k,
  \]
  which only requires squarings and products between matrices of the form \(A^{2^k}\).
    When \(A\) is Toeplitz-like the result is a straightforward consequence of \cref{prop:TxT};
when it is Hankel-like the product \(A^2\) is computed using \cref{prop:HxH}, the
remaining products are between Toeplitz-like matrices, and the result again follows from \cref{prop:TxT}.
%
%% By induction, the displacement rank of \(A^{2^i}\) is at most \(2^i\alpha+2^i-1\), and it takes
%%   \(\SO{n(\alpha2^i)^{\omega-1}}\) field operations to compute it from \(A^{2^{i-1}}\)
%%   by~\ref{prop:HxH} or~\ref{prop:TxT}.
%
 % The operations \(T^{2^{i+1}} =\left( T^{2^i}\right)^2\) and \(W_{i} = W_{i-1}
%%   \left(T^{2^i}\right)^{l_i}\) are both the product of two Toeplitz-like matrices of displacement rank
%%   at most \(2^{i} \alpha + 2^i - 1\). By \cref{prop:TxT}, they result in a matrices of displacement rank at most \(2^{i+1} \alpha + 2^{i+1} - 1\) and can be done in time
%%   \(\SO{n\left(\alpha 2^i\right)^{\omega-1}}\).
%%  The complexity of the whole step is :
%%   \begin{equation}
%%     \SO{n\alpha^{\omega - 1}\sum\limits_{i=0}^{\log r -1}\left( 2^{i(\omega - 1)}\right)}
%%     \end{equation}
\end{proof}
%%%%%
\subsection{Reconstruction of a Toeplitz+Hankel-like Matrix from its Generators}

The operator \(\nabla_{U_n,U_n}\) is defined in \cite[Section 4.5]{pan01} as partly-regular,
which means that a Toeplitz+Hankel-like matrix is completely defined by its generators \emph{and} its \emph{irregularity set} that contains all the entries in either its
first row, its last row, its first column or its last column.

A formula to recover a dense representation of the matrix from its generators and its first column is given in \cite[Theorem 4.5.1]{pan01}.

\begin{theorem}[{\cite{pan01}}]
    Let $M\in \K^{n \times n}$ be a Toeplitz+Hankel-like matrix, $G,H\in \K^{n \times \alpha}$ its generators and $c_0 = Me_0^n$ its first column, then
    \begin{align}
      M = \tau_{\UU{n}}(c_0) - \sum\limits_{j = 0}^{\alpha - 1} \tau_{\UU{n}}(G_{*,j})\trsp{\tau_{\Z{n}}(\Z{n}H_{*,j})}\label{eq:tph:slu}
    \end{align}
    where for an $n\times n$ matrix $A$ and a vector $v$ of length $n$ $\tau_A(v)$ denotes the matrix of the algebra generated by $A$ which has $v$ as its first column.
\end{theorem}

We show that one can derive a fast reconstruction algorithm for a Toeplitz+Hankel-like matrix from \cref{eq:tph:slu} and first detail
the structure of the various \(\tau_A(v)\) matrices.

\begin{lemma}
  $ \trsp{\tau_{\Z{n}}(v)}$ is the Toeplitz upper-triangular matrix with $\trsp v$ as its first row.
\end{lemma}

\begin{lemma}\label{lem:toq}
  $\tau_{\UU{n}}(v) = \sum\limits_{i = 0}^{n-1}v_iQ_i(\UU{n})$ where $Q_0(x) = 1$, $Q_1(x) = x$ and $Q_{i+1}(x) = xQ_i(x) - Q_{i - 1}(x)$.
\end{lemma}

\begin{proof}
The first column of \(Q_i(U_n)\) is \(e_i^n\).%\footnote{Vérifier que \(e_i\) est défini quelquepart}
\end{proof}

\begin{corollary}
  Column \(j\) of $\tau_{\UU{n}}(v)$ is $Q_j(\UU{n})v$.
\end{corollary}

\begin{proof}
  With \cref{lem:toq} and after checking the property for \(j\in\{0,1\}\), it suffices to prove \(Q_i(U_n)_{*,j+1}= \UU{n}Q_i(U_n)_{*,j} -Q_i(U_n)_{*,j-1}\). This is true for \(i\in\{0,1\}\) and if it is for $i$ and $i-1$, then
  \begin{align*}
    Q_{i+1}(U_n)_{*,j+1}&= \UU{n}^2Q_i(U_n)_{*,j} - \UU{n}Q_i(U_n)_{*,j-1}\\
    &-\UU{n}Q_{i-1}(\UU{n})_{*,j} + Q_{i-1}(\UU{n})_{*,i-1}
 % \todo[inline]{Il manque la preuve que \(Q_i(U_n)_{*,j} = Q_j(U_n)_{*,i}\)}
  %\begin{align*}
   % \tau_{\UU{n}}(v)_{*,j} &= \left(\sum\limits_{i = 0}^{n-1}v_iQ_i(\UU{n})\right)_{*,j}\\
%    &= \sum\limits_{i = 0}^{n-1}v_iQ_i(\UU{n})_{*,j}\\
   % &= \left(Q_0(\UU{n})_{*,j}\,|\,\cdots\,|\,Q_{n-1}(U_n)_{*,j}\right)v
  \end{align*}
\end{proof}

From these we can write the following proposition, inspired by \cite[Proposition 4.2]{heinig1988fast}, and which enables fast recursive reconstruction of the columns of a Toeplitz+Hankel-like matrix.
\begin{proposition}
\label{ppn:rec}
  
  Let $M\in \K^{n \times n}$ be a Toeplitz+Hankel-like matrix, $G,H\in \K^{n \times \alpha}$ its generators for \(\nabla_{U_n,U_n}\) and $c_0 = Me_0^n$ its first column. With the notation $c_{-1} = 0$, the columns $(c_k)_{0 \leq k \leq n-1}$ of $M$ follow the recursion:
  \begin{align}
    c_{k+1} = U_{n} c_k - c_{k-1} - \sum\limits_{j=0}^{\alpha - 1} H_{k,j}G_{*,j}\label{eq:rec}.
  \end{align}
  \end{proposition}

\begin{proof}
  Let $C$ be the matrix defined by the recursion formula and initial conditions of \cref{ppn:rec}, we will prove $C= M$.

By definition \(c_0\) is the first column of $M$; assume now that for \(j\leq k, c_j\) is column \(j\) of $M$, then \cref{eq:rec} can be detailed as
  \begin{align*}
    c_{k+1} %&= U_{n} c_k - c_{k-1} - \sum\limits_{j=1}^\alpha H_{k,j}G_{*,j}\\
%    &= U_{n} \left( Q_{k}(U_n)c_0 - \sum\limits_{j = 1}^\alpha \sum\limits_{i = 1}^k Q_{k}(U_n)G_{*,j}H_{i,j}\right)\nonumber\\
 %   &- \left( Q_{k-1}(U_n)c_0 - \sum\limits_{j = 1}^\alpha \sum\limits_{i = 1}^{k-1} Q_{k-1}(U_n)G_{*,j}H_{i,j}\right)\nonumber\\
  %  &- \sum\limits_{j=1}^\alpha H_{k,j}G_{*,j}\\
    &=  Q_{k+1}(U_n)c_0 - \sum\limits_{j = 0}^{\alpha - 1} \sum\limits_{i = 1}^{k-1} Q_{k+1}(U_n)G_{*,j}H_{i,j})\nonumber\\
    &- U_n\sum\limits_{j = 1}^\alpha Q_{k}(U_n)G_{*,j}H_{k,j} - \sum\limits_{j=1}^\alpha H_{k,j}G_{*,j} \\
%    &=  Q_{k+1}(U_n)c_0 - \sum\limits_{j = 1}^\alpha \sum\limits_{i = 1}^{k} Q_{k+1}(U_n)G_{*,j}H_{i,j}\\
    &= M_{*,k+1}\text{ by \cref{eq:tph:slu}}
  \end{align*}

  \end{proof}

\section{Material for Matrix Polynomials}\label{sec:matpoly}

We rely on the material from~\cite{KaVi05,Vil18}. For matrix polymonials and fractions the reader may refer to~\cite{Kailath80}. The rational matrix $H(x)=\trsp{V}M(x)^{-1}W$ over $\K(x)$
can be written as a fraction 
of two polynomial matrices. A right fraction description is given by square polynomial matrices $P(x)$ and $Q(x)$ such that 
$
H(x) = P(x) Q(x)^{-1} \in \K(x)^{m\times m},
$
and a left description by $P_l(x)$ and $Q_l(x)$ such that 
$
H(x) = Q_l(x)^{-1}R_l(x) \in \K(x)^{m\times m}.
$
Degrees of denominator matrices are minimized using column-reduced forms. A non-singular polynomial matrix 
is said to be column-reduced if its leading column coefficient matrix is non-singular~\cite[Sec.\,6.3]{Kailath80}.
We also have the notion of irreducible and minimal fraction descriptions. 
If $P$ and $Q$ (resp. $P_l$ and $Q_l$) have unimodular right (resp. left) matrix gcd's~\cite[Sec.\,6.3]{Kailath80}
then the description is called irreducible.  If $Q$ (resp. $Q_l$) is column-reduced then the description is called 
minimal.

For a given $m$, define $1 \leq \nu \leq n$ to be the sum of the degrees of the first $m$ largest invariant factors of 
$M(x)$ (equivalently, the first $m$ diagonal elements of its Smith normal form).
The following will ensure that the minimal polynomial of $T$, which is the largest invariant factor of $M(x)$ can be computed from the Smith normal form of an appropriate denominator $Q(x)$; 
see \cref{cor:bsgs}.

\begin{theorem}  (\cite[Thm. 2.12]{KaVi05} and \cite{Vil97:TR}) \label{twosidedproj}
  Let \(V\) and \(W\) be block vectors over a sufficiently large field \(\K\) whose
  entries are sampled uniformly and independently from a finite subset \(S \subseteq \K\).
  Then with probability at least \(1-2n/|S|\), $H(x)=\trsp{V}M(x)^{{-1}}W$ has left and right irreducible descriptions with denominators of 
  degree $\lceil \nu /m\rceil$, of determinantal degree $\nu$, and whose $i^\text{th}$ invariant factor (starting from the largest degree) is the $i^\text{th}$ invariant factor of~$M(x)$. 
\end{theorem}

The next result we need is concerned with the computation of an appropriate denominator $Q$ as soon as the truncated
power series in \cref{eq:defseries} is known. We notice that $H(x)=\trsp{V}M(x)^{-1}W$ is strictly proper in that it tends to zero when \(x\) tends to infinity.
For fraction reconstruction we use the computation of minimal approximant bases (or $\sigma$-bases)~\cite{BeLa94,BarBul92}, and the algorithm with complexity bound $\SO{m^{\omega-1}n}$ in~\cite{GJV03,JeannerodNeigerVillard2020}.

\begin{theorem}[{\cite[Lemma\,3.7]{GJV03}}] \label{lem:minapprox} 
Let $H \in \K(x)^{m\times m}$ be a strictly proper power series, with left and right matrix fractions descriptions of degree at most $d$. A denominator $Q$ of a right irreducible description 
$H(x)=P(x)Q(x)^{-1}$ can be computed in $\SO{m^{\omega-1}n}$ arithmetic operations from the first 
$2d+1$ terms of the expansion of $H$.   
\end{theorem}

In our case, from \cref{twosidedproj} we will obtain the existence of appropriate fractions of degree less than $\lceil n/m\rceil$, and use \cref{lem:minapprox} for bounding the cost of the computation of $Q$.

%%% Local Variables:
%%% TeX-master: "charpoly-sigmalu.tex"
%%% End:

%% file: bsgs.tex
%!TEX root = charpoly-sigmalu.tex

%%%%%%%%%%%%%%

\section{A Baby-Step Giant Step algorithm}
\label{sec:bsgs}

% In this section, we consider classes of structured matrix either stable by multiplication or the
%% class of Hankel-like matrices, having generator representation with respect to a regular
%% displacement operator.
%% This includes Toeplitz-like, Hankel-like, Vandermonde-like matrices. \todo{anything else?}

%% %% In this section, we will only consider an input matrix \(T\) with the  Toeplitz-like structure.
%% Such matrices have an image with rank \(\alpha\) through the Stein displacement operator
%% \(\Delta_{Z,\trsp{Z}}\). They can therefore be represented with generators \(G,H\in
%% \K^{n \times \alpha}\): \(\Delta_{Z,\trsp{Z}}(T)=G\trsp{H}\).

In this section, we propose a direct adaptation of the baby steps/giant steps variant of Coppersmith's
block-Wiedemann algorithm from~\cite[Sec.\,4]{KaVi05} to the case of structured matrices.
In order to compute the terms of the series~\eqref{eq:defseries}, we will assume that the input matrix
\(T\) has been inverted, using~\cite[Theorem~6.6]{BJMS17}. In this section we will therefore denote
by~\(T\) this inverse and compute the projections of its powers.
\subsection{Description of the Algorithm}
Let $V,W \in \K^{n\times m}$ be the block vectors used for the projection. 
\cref{alg:bsgs} performs $r$ baby steps and $s$ giant steps to compute the first terms of the sequence
\(H_k= \trsp{V} T^k W=\trsp{V} (T^r)^jT^i W\) for \(0\leq k \leq 2\lceil n/m\rceil\), \(0\leq i<r\), \(0\leq j < s\) and \(rs \geq k +1\).

%\footnote{ We follow the lines of  \cite[Sec.\,4]{KaVi05}.}

\begin{algorithm}[htb]
\caption{Compute $H_k= \trsp{V} T^k W$ for $0\leq k \leq 2\lceil n/m\rceil$}\label{alg:bsgs}
\begin{algorithmic}[1]
  \Require {Generators of \(T \in \mathbb{K}^{n\times n}\), Toeplitz-like or Hankel-like}
  \Require {\(m,r,s\in\N \text{ s.t. }rs\geq 2\lceil n/m\rceil +1\)}, \(r\) even if \(T\) is Hankel-like
  \Require {\(V,W \in \K^{n\times m}\)}
\Ensure{\(H = \left(H_{rj+i}\right)_{j<s,i<r}\) where \(H_k = \trsp{V}T^kW\)}
\State \(W_0 \assign W\)
\For{\(1\leq i \leq r-1\)}
\State \(W_i \assign TW_{i-1}\)\label{step:bsgs:bs}
\EndFor
\State \(R \assign T^r \) \label{step:bsgs:gsinit}
\State \(V_0 \assign  V\)
\For{\(1\leq j \leq s-1\)}
\State \(\trsp{V_j} \assign \trsp{V_{j-1}}R\) \label{step:bsgs:gs}
\EndFor
\State \(H \assign  %\left(\begin{array}{c} V_0 \\ \vdots\\ V_{s-1} \end {array}\right) 
\trsp{
\begin{pmatrix}
  {V_0} &\dots& {V_{s-1}}
\end{pmatrix}}
\begin{pmatrix}
  W_0 &\dots& W_{r-1}
\end{pmatrix}
\) \label{step:bsgs:prod}
\end{algorithmic}
\end{algorithm}

This algorithm relies on three main operations:
\begin{enumerate}
\item the product  of a structured matrix to dense rectangular matrix, supported by~\cref{prop:mbv}
  for~\cref{step:bsgs:bs,step:bsgs:gs};
\item the exponentiation of a structured matrix, supported by~\cref{prop:power} for~\cref{step:bsgs:gsinit};
\item the product of two dense rectangular matrices for~\cref{step:bsgs:prod}.
\end{enumerate}
%% \begin{proof}[Correctness]
%%   Since for all \(0\leq i<r, Y_{i} = T^i Y\) and for all
%%   \(0\leq s, X_j = \trsp{X} \left(T^r\right)^j\)
%%   we have
%%   \[X_jY_i = \trsp X T^{rj+i} Y = H_{rj+i}.\]
%% \end{proof}

%%%%%%%%%%%%%%%%%%%%%%%%%%%%%%%%%
\subsection{Cost Analysis}

%% The following Lemma gives a bound on the complexity for the initialisation of the giant steps in \cref{step:bsgs:gsinit}.
%% \begin{lemma}\label{lem:gsinit}
%%   Computing \(T^r\) can be done in \(\SO{n(\alpha r)^{\omega - 1}}\) operations in \(\K\).
%% \end{lemma}

%%   \begin{proof}
%% o
%%   \end{proof}
  %%%%%%%%%%%
\begin{theorem}\label{thm:bsgs}
  \cref{alg:bsgs} runs in
  \(
  \SO{\bsgscost}
%  \SO{n^{\frac{\omega^2-4\omega-1}{\omega-5}}\alpha^{\frac{\omega-1}{5-\omega}}}
  \) operations in~\K for well chosen \(m\), \(r\) and \(s\).
\end{theorem}

For instance, when the displacement rank \(\alpha\) is constant, and with the best known estimate
\(\omega = 2.373\) \cite{alman2020refined} the cost becomes  \( \SO{n^{1.851}}\) while
  it is \(\SO{n^2}\) for \(\omega=3\).

\begin{proof}
%  We will assume \(\alpha \leq m\).\todo[inline]{to be justified}
  From~\cref{prop:mbv}, applying an \(n\times m\)
  block to \(T\) can be done in \SO{n\max(m,\alpha) \min(m,\alpha)^{\omega-2} } field operations. Hence the \(r\) baby-steps,
  \cref{step:bsgs:bs}, computing the \((T^i W)_{0\leq i < r}\) cost overall
  \begin{equation}\label{eq:bs}
    \SO{nr\max(m,\alpha) \min(m,\alpha)^{\omega-2} }
    %=\SO{n^{1+\mu+\rho}\alpha^{\omega-2}}
  \end{equation}
  field operations.%\footnote{Rectangular product could also be used here}

By~\cref{prop:power}, the initialization of the giant steps, \cref{step:bsgs:gsinit} computing a
structured representation for \(T^r\), can be done in
\begin{equation}\label{eq:gsinit}
  \SO{nr^{\omega - 1}\alpha^{\omega - 1} }
  %=\SO{n^{1+\rho(\omega-1)}\alpha^{\omega-1}}
\end{equation}
operations in $\K$. 

  Then each of the giant steps, \cref{step:bsgs:gs}, is a product of an \(m\times n\) dense matrix with
  an \(n\times n\) matrix of displacement rank \(\alpha r\).  From~\cref{prop:mbv}, these
  \(s\) steps cost
  \begin{equation}\label{eq:gs}
    \SO{ns\max(m,\alpha r) \min(m,\alpha r)^{\omega-2} }
  \end{equation}

  Lastly, the computation of the product resulting in \(H\), \cref{step:bsgs:prod}, uses
  \(\SO{n \max(mr, ms) \min(mr,ms)^{\omega - 2}}\) or equivalently 
  \begin{equation}\label{eq:prod}
    \SO{nm^{\omega-1}\max(r,s)\min(r,s)^{\omega-2}}
%    \SO{n^{1+\mu(\omega-1) + \max(\rho,\sigma)+\min(\rho,\sigma)(\omega-2)}}
  \end{equation}
  field operations.

  Let \(m =\left\lceil n^{\frac{\omega-3}{\omega-5}} \alpha^{\frac{2}{5-\omega}}\right\rceil\) and set
  \(r=s=\left\lceil\sqrt{2n/m}\right\rceil\).
  Note that \(\alpha\leq m\leq \alpha r\).
  Therefore~\eqref{eq:bs} is dominated by~\eqref{eq:prod}.
  Moreover~\eqref{eq:gs} writes
  \(
  \SO{n^2m^{\omega-3}\alpha}, 
  \)
  \eqref{eq:prod} writes
  \(
  \SO{n^{\frac{\omega+1}{2}}m^{\frac{\omega-1}{2}}}
  \)
  and both terms equal
  \[
  \SO{\bsgscost}.
%  \SO{n^{\frac{\omega^2-4\omega-1}{\omega-5}}\alpha^{\frac{\omega-1}{5-\omega}}}.
  \]
  Finally,~\eqref{eq:gsinit} writes
  \(\SO{n^{\frac{\omega+1}{2}}(\frac{\alpha^2}{m})^{\frac{\omega-1}{2}}}\)
and is thus  dominated by~\eqref{eq:prod}.
  \end{proof}

Let us now suppose that the entries of \(V\) and \(W\) are sampled uniformly and independently from a finite subset \(S\subseteq \K\), we then have the following:
\begin{corollary}\label{cor:bsgs}
  The  minimal polynomial of an \(n\times n\) Toeplitz-like or Hankel-like matrix with displacement rank 
  \(\alpha\) can be computed by a Monte Carlo algorithm   
  in
  \[
  \SO{\bsgscost}
%  \SO{n^{\frac{\omega^2-4\omega-1}{\omega-5}}\alpha^{(\omega-1)/(5-\omega)}}
  \]
  field operations with a probability of success of at least \(1-(n^2+3n)/|S|\).
\end{corollary}

\begin{proof}
  The first step is to compute the inverse of \(T\), using~\cite[Theorem~6.6]{BJMS17} in
  \SO{n\alpha^{\omega-1}} operations in \K. Then running~\cref{alg:bsgs} on \(T^{-1}\) costs
  \(\SO{\bsgscost}\) which dominates  \SO{n\alpha^{\omega-1}} since \(\alpha\leq n\).
  From the sequence of matrices \((H_k)_{0\leq k\leq 2n/m}\), one can compute a minimal denominator \(Q\)
  for $H(x)=\trsp{V}(x\I{n}-T)^{-1}W \in \K[x]^{m\times m}$ in \(\SO{nm^{\omega-1}}\) field operations, by~\cref{lem:minapprox}.

Using \cref{twosidedproj}, the minimal polynomial is then obtained  as the first invariant factor in the Smith form of \(Q\),
  computed by~\cite[Proposition~41]{Sto03}. This step also costs  \(\SO{nm^{\omega-1}}\) field
  operations and
%  Since \(\mu<1\), we have
%  \[
%  1+(\omega-1)\mu < \frac{\omega+1}{2}+\frac{\omega-1}{2}\mu
%  \] which
  since \(m\leq n\) we have
  \[
      nm^{\omega - 1} \leq n^\frac{\omega+1}{2}m^\frac{\omega-1}{2}
  \] which
  shows that the cost of these last two computations will always be dominated by the cost of the
  product~\eqref{eq:prod}. 
The probability of failure for the computation of \(T^{-1}\) is \(n(n+1)/|S|\) by~\cite[Lemma~6.2]{BJMS17}.
A union bound combining this probability and the failure probablity of~\cref{twosidedproj} yields
a probability of failure of \((n^2+3n)/|S|\).
\end{proof}

Note that this result carries over to the computation of the characteristic polynomial of any
Toeplitz-like or Hankel-like matrix \(T\) having fewer than \(m\) invariant factors in its Frobenius normal form.

%% file: sigmalu.tex
\section{An algorithm based on structured inversion}
%\section{Using compressed representations}
\label{sec:compressed}

In this section we propose an algorithm computing the determinant of a generic structured polynomial
matrix~\(M\in\K[x]^{n\times n}\) with displacement rank \(\alpha\)
based on the structure of the \SLU representation of
Toeplitz-like matrix, or a generalization thereof for Hankel-like matrices, as presented in~\eqref{eq:tph:slu}.

\begin{paragraph}{Principle of the algorithm}
Here, the sequence \((H_k = \trsp{V}T^{-k} W)_k\) is obtained as the matrix coefficients of the series
expansion \(\trsp{V}M^{-1}W\). As \(2\lceil n/m\rceil +1\) terms are required, and with the special choice
    \(V = W = X=\trsp {\begin{pmatrix}  I_m&|& 0\end{pmatrix}}\), this boils down to computing a dense representation of
the \(m\times m\) leading principal submatrix of \(M^{-1}\mod x^{2\lceil n/m\rceil+1}\).
The outline of the algorithm is as follows.
\begin{enumerate}
\item Compute the inverse \(M^{-1}\mod x^{2\lceil n/m\rceil+1}\) in a compressed representation
\item Crop this representation to form a representation of the \(m\times m\) leading principal
  submatrix;
\item Extract the dense representation from this representation.
\end{enumerate}
We will now present the algorithm specialized for the two classes of interest.
\end{paragraph}

\subsection{The Algorithm for Toeplitz-like Matrices}

A Toeplitz-like matrix \(T\) is represented by a pair of generators \(G,H \in \K^{n\times\alpha}\)  satisfying
\(T=\sum_{i=0}^{\alpha-1}L(G_{*,i})L(H_{*,i})^T\), where \(L(v)\) is the lower triangular Toeplitz
matrix with \(v\) as its first column \cite{KKM79, Kal94}.
The \(m\times m\) leading principal submatrix of any product \(L(v)\trsp{L(w)}\) is the product of
the \(m\times m\) leading principal submatrix of these factors, which in turn is
\(L(v_{1..m})\trsp{L(w_{1..m})}\). 
\cref{alg:slu:toep} relies on this property to produce \(S^{(m)}\) from the \(m\) first rows of
the generators of \(T^{-1}\).

\begin{algorithm}
  \caption{Compute \(S^{(m)}\): Toeplitz-like case
    %Compute a dense representation of $S^{(m)}= \trsp{X} M^{-1} X\mod x^{2\frac{n}{m}+1}$ for
    %$X=\left( I_m ~~  0\right)$ } %$
  }
  \label{alg:slu:toep}
\begin{algorithmic}[1]
\Require{\((G, H)\) generators of \(M\in \K[x]^{n\times n}\), a Toeplitz-like matrix of displacement rank \(\alpha\)}
\Ensure{Dense representation of \(S^{(m)}= \trsp{X} M^{-1} X \bmod x^{2\lceil n/m \rceil +1}\)}
\State \((E,F)\assign\) generators for \(M^{-1}\bmod x^{2\lceil n/m \rceil +1}\) \label{step:slu:inv}
\State \(E'\assign \trsp{X}E\); \(F'\assign FX\)
\State \(S^{(m)}\assign\sum_{i=0}^{\alpha-1}L(E'_{*,i})\trsp{L(F'_{*,i})} \mod x^{2\lceil n/m \rceil +1}\)  \label{step:slu:exp}
\end{algorithmic}
\end{algorithm}

%% Let $M \in \K[x]^{n\times n}$ be a Toeplitz-like polynomial matrix of degree $1$, and let $\phi:\K[x]^{n\times n}\rightarrow\K[x]^{n\times n}$ be a corresponding displacement operator. We assume that $M$ has constant displacement rank $\alpha$.\footnote{$M$ is not necesarily the characteristic matrix of $T$}   \footnote{The determinant of $M$ has degree at  $n$, it can be conputed in $n^{2+o(1)}$ operations, 
%% see for instance~\cite{Pan92}. Here we show it can be done in less operations.}

%where the $L_i$'s and the $U_i$'s are lower- and upper-triangular Toeplitz matrices.
%% The $m\times m$ leading principal submatrix $S^{(m)}$ of $S$ satisfies 
%% \begin{align}
%% S^{(m)}(x) &= \sum _{i=1}^{\alpha} L_i^{(m)}(x) U_i^{(m)}(x) ~\bmod x^{2n/m}\\ &\in \K[x]^{m\times m}
%% \end{align}
%% where the $L_i^{(m)}$'s and the $U_i^{(m)}$'s are themselves the Toeplitz leading principal submatrices 
%% of the $L_i$'s and the $U_i$'s.

%\subsection{Correctness and cost analysis}

\begin{theorem}\label{th:slu}
  \cref{alg:slu:toep} is correct for \(M=x\I{n}-T\) and \(T\) generic and uses
  \[\SO{\frac{n^2}{m}\alpha^{\omega-1}+nm\alpha}\] operations in \K.
\end{theorem}

\begin{proof}
  From the above remark, \(E'=E_{1..m,*}\) and \(F'=F_{1..m,*}\) are generators for \(S^{(m)} =
\trsp{X}M^{-1}X\).
  Note that no division by \(x\) in the  ring \(\K[x] / \langle x^{2\lceil n/m \rceil +1} \rangle\) will occur
  in~\cref{step:slu:inv} as \(T\) has generic rank   profile, and consequently all leading principal
  minors of \(M(x)\) are not divisible by \(x\) which shows the correctness.
  
By~\cite[Theorem~34]{BJMS17}, \cref{step:slu:inv}, computing the generators of $M^{-1}$,  can be
computed in \(\SO{n\alpha^{\omega-1}}\) operations over \(\K[x]/\langle x^{2\lceil n/m \rceil +1}\rangle\) which in turn is
    \begin{equation}\label{tl:inv}
%\(
        \SO{\frac{n^2}{m}\alpha^{\omega-1}}
% \)
\end{equation}
operations    in \(\K\).
%and truncation to the $m\times m$ leading principal submatrix is free.

The dense reconstruction of \(S^{(m)}\) in~\cref{step:slu:exp} is achieved by \(\alpha\) products of
an \(m\times m\) Toeplitz matrix \(L(E'_{*,i})\) by an \(m\times m\) dense matrix
\(\trsp{L(F'_{*,i})}\) for a total cost of 
    \begin{equation}\label{tl:rec}
%\(
        \SO{nm\alpha}
% \)
\end{equation}
operations    in \(\K\).
\end{proof}

\begin{corollary}\label{cor:slu}
  The characteristic polynomial of a generic \(n\times n\) Toeplitz-like matrix with displacement rank
  \(\alpha\) can be computed in
   \(
 \SO{n^{2 - \frac{1}{\omega}}\alpha^{\frac{(\omega-1)^2}{\omega}}}
 \)
 operations in \K when \(\alpha   = O\left(n^\frac{\omega -2}{-\omega^2 + 4\omega - 2}\right)\), 
 and $\SO{n^\frac{3}{2}\alpha^\frac{\omega}{2}}$ 
 otherwise. 
\end{corollary}

Note that this is $\GO{n^{1.579}}$ (resp. $\GO{n^{1.667}}$) for $\alpha$ constant and \(\omega
= 2.373\) (resp. $\omega = 3$). When \(\alpha = \Theta\left( n^\frac{\omega -2}{-\omega^2 + 4\omega - 2}\right)\) and taking \(\omega = 2.373\) (resp. \(\omega = 3\)),
both expressions become \SO{n^{1.74}} (resp. \SO{n^{3}}).
%  When \(\alpha\) is constant and for \(\omega = 2.3728639\), this is \(\SO{n^{1.58}}\) while
%  it is \(\SO{n^{1.67}}\) for \(\omega=3\).\footnote{This is \cite[Thm. 1.1]{Vil18} with $d=1$.}

The complexity when \(\alpha\) is low can also be written as \[\SO{n^{\omega -
    f(\omega)}\alpha^{f(\omega)}},\]
similarly as in~\cref{thm:bsgs}, which can be interpreted as a transfer of part of the exponent from $n$ to $\alpha$ by using the
structure of the matrix.
\begin{proof}

  The family of Toeplitz matrices presented in~\cref{sec:point:toep} proves that for a generic
  Toeplitz-like matrix \(T\), the matrix
  \(\mathcal H^{(n)}=\mathcal{H}_{1..n,1..n}\) is non-singular, where
$$\mathcal H = \left(\trsp VT^{i+j}W\right)_{0\leq i,j \leq \lceil n/m\rceil - 1}.$$ Then \cite[Lemma~2.4]{Vil18} implies that the irreducible
  left and right fractions descriptions of \(\trsp{X}M^{-1}X\) have degree at most \(\lceil n/m \rceil \).
  Thus \cref{lem:minapprox} ensures that an appropriate denominator $Q$ of a right fraction description of
\(\trsp{X}M^{-1}X\) can be computed from \(S^{(m)}=\trsp{X}M^{-1}X \mod x^{2\lceil n/m \rceil +1}\).

  Besides the computation of \(S^{(m)}\) by~\cref{th:slu}, the computation of the denominator $Q$ of its
  irreducible right fraction description costs
%  \(
  \begin{equation}\label{tl:det}
        \SO{nm^{\omega-1}}
 %       \)
\end{equation}
operations by~\cref{lem:minapprox}. Computing the determinant of~\(Q\) has same cost by~\cite{Sto03,GuSt12}.
The total cost depends on \(\alpha\).
\paragraph{Case 1: $\alpha = \GO{n^\frac{\omega -2}{-\omega^2 + 4\omega - 2}}$.}
    We set $m = n^\frac{1}{\omega}\alpha^\frac{\omega-1}{\omega}$
    so that \(\alpha = \GO{m^{\omega-2}}\) and
    the term \eqref{tl:rec} is dominated by \eqref{tl:det}. For the chosen
    value of $m$ the terms \eqref{tl:inv} (decreasing in $m$)
    and \eqref{tl:det} (increasing in $m$) are equal, leading to
    a full cost of
    \SO{n^{2 - \frac{1}{\omega}}\alpha^{\frac{(\omega-1)^2}{\omega}}} operations in \(\K\).

\paragraph{Case 2: $\alpha = \GOM{n^\frac{\omega -2}{-\omega^2 + 4\omega - 2}}$.}
We  set $m = n^\frac{1}{2}\alpha^\frac{\omega-2}{2}$ so that
    $\alpha = \GOM{m^{\omega-2}}$.
    In this case the term \eqref{tl:det} is dominated by \eqref{tl:rec}
    and for this value of $m$ we have equality between the terms \eqref{tl:inv} and \eqref{tl:rec}, leading to
    a full cost of $\SO{n^\frac{3}{2}\alpha^{\frac{\omega}{2}}}$ operations in \(\K\)
\kern-0.35em. %Mettre le point à la ligne du dessus décalait le \qed
\end{proof}

\subsection{The Algorithm for Hankel-like Matrices}

In this section we are interested in adapting the previous algorithm to Hankel-like matrices. If $T$ is Hankel-like then $M(x) = x\I{n} - T$ is Toeplitz+Hankel-like.

We will thus generalize and consider that $T$ is a Toeplitz+ Hankel-like matrix. We are interested in computing the first \(2\lceil{n}/{m}\rceil+1\) terms of the series $\trsp XM(x)^{-1}X$. We are going to adapt the Toeplitz algorithm and use Pan's Divide-and-Conquer algorithm for inversion \cite[Chapter 5]{pan01}. Computing the characteristic polynomial from there does not depend on the structure of $M$ or $T$. 
                                           
The strategy consists in computing generators for the truncated matrix from which we can recover a dense representation. Algorithm \ref{alg:sm} details the steps. The generators and irregularity set of the inverse in \cref{step:sm:inv} are computed with Pan's Divide and Conquer algorithm \cite{pan01}, as well as the solution to the linear system. The following lines are dedicated to the reconstruction of the dense representation of $S^{(m)}(x)$  from the generators. The correctness of \cref{alg:sm} is proved by Proposition \ref{ppn:rec}.
\begin{algorithm}
  \caption{Compute \(S^{(m)}\): Toeplitz+Hankel-like case
    %    Compute a dense representation of $S^{(m)}(x)= \trsp X M^{-1}(x) X \mod x^{2\frac{n}{m}+1}$
  }
  \label{alg:sm}
\begin{algorithmic}[1]
\Require $(G, H, v)$ generators and irregularity set of $M \in \mathbb{K}[x]^{n\times n}$, a
Toeplitz+Hankel-like matrix of displacement rank \(\alpha\).
\Ensure Dense representation of $S^{(m)}(x)= \trsp X M^{-1}(x) X \mod x^{2\lceil{n}/{m}\rceil+1}$
\State $(E,F, u), c \assign $ generators and irregularity set of the inverse of $M$, solution of \(Mc = e_0^n\) \label{step:sm:inv}
\State $c_0 \assign \trsp Xc$ \label{step:sm:init}
\State $c_1 \assign \UU{m} c_0 - \sum\limits_{i = 0}^{\alpha - 1}E_{0,i}F_{0\hdots m-1,i}$
\For{$1\leq k\leq m - 2$}
\State $c_{k+1} \assign \UU{m} c_k - c_{k - 1} - \sum\limits_{i = 0}^{\alpha - 1}E_{k,i}F_{0\hdots m-1,i}$ \label{step:sm:end}
\EndFor
\State $S^{(m)}(x) = (c_0||\cdots||c_{m-1})$ 
\end{algorithmic}
\end{algorithm}

\begin{theorem}\label{thm:slu:tph}
  ~\cref{alg:sm} is correct for \(M=x\I{n}-T\) and \(T\) generic and uses
  \[
  \SO{\frac {n^2}{m}  \alpha^2 + mn\alpha}
  \]
  operations in \K.
    \end{theorem}

\begin{proof}
\cref{step:sm:inv} can be done in $\tilde O (\alpha^2n)$ operations in the base ring, so $\tilde O\left(\frac {n^2} m \alpha^2 \right)$ operations on $\K$ \cite[Corollary 5.3.3]{pan01}.
Each step of the \texttt{for} loop consists of a number of polynomial operations modulo $x^{2 \lceil n/m \rceil + 1}$ linear in $m\alpha$ as $\UU{m}$ has only two non-zero entries on each row. Lines \ref{step:sm:init} to \ref{step:sm:end} can be done in $\tilde O (m^2\alpha)$ operations in the base ring, so $\tilde O({n} m\alpha )$ operations on $\K$.
\end{proof}

The minimal polynomial is then obtained the same way as in \cref{sec:bsgs} which leads to~\cref{cor:slu:hank}.

\begin{corollary} \label{cor:slu:hank}
  The  characteristic polynomial of a generic \(n\times n\) Toeplitz+Hankel-like matrix with
  displacement rank  \(\alpha\)
  %, generic rank profile and fewer than $m$ invariant factors in its Frobenius normal form
  can be computed in
      \SO{n^{2 - \frac{1}{\omega}}\alpha^{\frac{2(\omega-1)}{\omega}}}
  field
    operations when $\alpha = \GO{n^\frac{\omega -2}{4-\omega}}$, and $\SO{n^\frac{3}{2}\alpha^\frac{3}{2}}$ otherwise. 
 \end{corollary}

 The complexity in $n$ is the same as in the Toeplitz-like case but there is a stronger dependence in $\alpha$ as there is no known algorithm to compute the inverse of a Toeplitz+Hankel-like matrix in $O(n\alpha^{\omega - 1})$, the best one depending on $\alpha^2$.
 
\begin{proof}
  The family of Hankel matrices presented
  in~\cref{sec:point:hank} now proves that for all generic Hankel-like matrix \(T\), the matrix
  \(\mathcal H^{(n)}\) is
  non-singular.
  The rest of the proof is similar to the Toeplitz-like case in~\cref{cor:slu}.
  %% Then \cite[Lemma~2.4]{Vil18} implies that the irreducible
  %% left and right fractions descriptions of \(\trsp{X}M^{-1}X\) have degree at most \(\lceil n/m \rceil \).
  %% Thus \cref{lem:minapprox} ensures that the denominator \(G\) of the right fraction description of
  %% this power series can be computed
  %% %in \SO{m^{\omega-1}n}
  %% from the first \(2\lceil n/m \rceil +1\) terms of the sequence
  %% \(H_k\), or equivalently, from \(S^{(m)}=\trsp{X}M^{-1}X \mod x^{2\lceil n/m \rceil +1}\).

  Again the overall cost is that for computing the denominator and its determinant in
  \(\SO{nm^{\omega-1}}\) operations in \K plus the cost of computing the sequence $H_k$. We
  distinguish two cases:
\paragraph{If $\alpha = \GO{n^\frac{\omega -2}{ 4- \omega }}$:}
Setting  $m = n^\frac{1}{\omega}\alpha^\frac{2}{\omega}$
so that \(\alpha = \GO{m^{\omega-2}}\) and the full cost is
    \SO{n^{2 - \frac{1}{\omega}}\alpha^{\frac{2(\omega-1)}{\omega}}}. 

\paragraph{If $\alpha = \GOM{n^\frac{\omega -2}{ 4-\omega }}$:}
Setting $m = n^\frac{1}{2}\alpha^\frac{1}{2}$ so that
    $\alpha = \GOM{m^{\omega-2}}$ and the full cost is
 $\SO{n^\frac{3}{2}\alpha^{\frac{3}{2}}}$.
\end{proof}

%%% Local Variables:
%%% TeX-master: "charpoly-sigmalu.tex"
%%% End: